\documentclass{biometrika}
\usepackage{fancyhdr} 
\usepackage{amsmath}
\usepackage{bm}
\usepackage{graphicx}
\usepackage{times}
\usepackage[linesnumbered,ruled,vlined]{algorithm2e}
\usepackage{algorithmic}
\usepackage{mathrsfs}
\usepackage{amssymb}
\RequirePackage{natbib}
\makeatletter
\renewcommand{\algocf@captiontext}[2]{#1\algocf@typo. \AlCapFnt{}#2} 
\def\@algocf@capt@plain{top}
\renewcommand{\algocf@makecaption}[2]{%
  \addtolength{\hsize}{\algomargin}%
  \sbox\@tempboxa{\algocf@captiontext{#1}{#2}}%
  \ifdim\wd\@tempboxa >\hsize
    \hskip .5\algomargin%
    \parbox[t]{\hsize}{\algocf@captiontext{#1}{#2}}
  \else%
    \global\@minipagefalse%
    \hbox to\hsize{\box\@tempboxa}
  \fi%
  \addtolength{\hsize}{-\algomargin}%
}
\makeatother
\begin{document}

\title{Mixed and  missing data: a unified treatment with latent graphical models}
\date{}

\author{Xiao Li, Jinzhu Jia and Yuan Yao}
\affil{LMAM, School of Mathematical Sciences, and Center for Statistical Science,\\ Peking University, Peking, China 100871 \email{seanli@pku.edu.cn,\{jzjia,yuany\}@math.pku.edu.cn}}

\maketitle

\begin{abstract}
We propose to learn latent graphical models when  data have mixed variables and missing values. This model could be used for further data analysis, including regression, classification, ranking etc. It also could be used for imputing missing values. 

We specify a latent Gaussian model for the data, where the categorical variables are generated by discretizing an unobserved variable and the latent variables are multivariate Gaussian. The observed data consists of two parts: observed Gaussian variables and observed categorical variables, where the latter part is considered as partially missing Gaussian variables. 
We use the Expectation-Maximization algorithm to fit the model.  To prevent overfitting we use sparse inverse covariance estimation to obtain sparse estimate of the latent covariance matrix, equivalently, the graphical model. The fitted model then could be used for problems including regression, classification and ranking. Such an approach is applied to a medical data set where our method outperforms the state-of-the-art methods.   Simulation studies and real data results suggest that our proposed model performs better than random forest in terms of prediction error when the model is correctly specified, and is a better imputation method than hot deck imputation even if the model is not correctly specified.
\end{abstract}


\section{Introduction}

The problem of missing data arises in many situations, especially in clinical trials and social investigations. In general settings, the explanatory variables contain both continuous variables and categorical variables (we refer this kind of data as mixed data). As such, a major difficulty of this kind of problems stems from modeling the data. It is often very hard to specify a unified model for the variables, let alone making inference and imputing missing values.

Let's look at the PCI (percutaneous coronary intervention) data provided by two hospitals in Beijing, China, which has been studied by  \cite{wang2015risk}  from a clinical perspective.  The data consists of 38 variables that are measured before one patient's heart operation and a binary variable indicating whether there is reflux after PCI. There are totally 2582 patients collected. The objective of the analysis is to determine which ones, among the 38 variables, are related to the outcome of the operation and to predict the outcome of the operation. 23 of the 39 variables are categorical variables with the levels ranging from two to six. In addition, all 39 variables have missing values and the missing proportion ranges from less than $1\%$ to more than $70\%$.  To analyze this data, it is crucial 
to propose appropriate methods for mixed and missing data.

In many cases including the PCI data describe above, it is reasonable to assume that the categorical variables and rank variables are formed by discretizing a latent variable \citep{skrondal2007latent}. For instance, in medical science, the latent variables can represent physiological conditions that exist in hidden form but are not directly measurable, and instead, they can be measured indirectly by some surrogate variables, like the outcome of an operation. Under the latent variable assumption, the latent Gaussian copula model was considered by \cite{xue2012nonpara} and \cite{fan2014latent}. This semiparametric model has many nice theoretical properties, but is not suitable for inference with missing data. The model reduces to the latent Gaussian model when the latent variables are multivariate Gaussian. \cite{han2012composite}  use composite marginal likelihood function to estimate the latent parameters for the latent Gaussian model. To the best of our knowledge, there has not been any likelihood-based method for inference under the latent Gaussian model when there are missing values in the observed data. 

We propose a likelihood based method along with the Expectation-Maximization algorithm to estimate the latent parameters. The fitted model is then used for classification. It also could be used for regression and ranking without any further difficulty than classification. The value of our proposed method is twofold. (1) It provides new tracks for data analysis with mixed data, whether missingness exists or not. Numerical results show that our classification model performs better than random forest in terms of prediction error, and that the latent Gaussian model is a better imputation model than hot deck imputation \citep{tanner1991tools}. (2) Our method serves as a supplement to the estimation of Gaussian graphical model with mixed and missing data. For a review of related literature on estimation of Gaussian graphical models, see \cite{friedman2008glasso}, \cite{banerjee2008binary}, \cite{cai2011clime},\cite{xue2012ising}, \cite{lee2012mixed}, \cite{fel2013rf} and \cite{buhlmann2012missing}. Simulations show that the inverse covariance matrix estimated by our method is close to the true covariance matrix even when missing values exist.





The rest of the paper is organized as follows. In Section 2 we introduce basic notations and formaly describe our model for mixed and missing data. In Section 3 we introduce the likelihood-based method to find the maximum likelihood estimation of parameters in our model. In Section 4, we discuss classification methods for mixed and missing data. Numerical results are given in section 5. \par

\section{Notations and Model Descriptions}
Let $X\in \mathbb R^{n\times p}$ be our collected data that consist both continuous and discrete variables. We view  rows of $X$ as independent and identically distributed observations.  Each column of $X$ corresponds to one random variable. We use $X_j$ to denote one random variable and the $j$th column of $X$ interchangeably. Suppose there are $p_1$ continuous variables and $p_2(=p-p_1)$ categorical variables. For a categorical variable $X_j$, let $n_j$ be the levels of $X_j$. Without loss of generality, suppose that $X_j$ takes value from $0,\dots, n_j-1$. For the $i$th row in $X$, denote $X_{i,con}$, $X_{i,cat}$  as the observations of the continuous and categorical variables respectively. In general we assume that $X$ contains missing data that are missing at random. Denote $X_{obs}$ as the collection of all observed data. Specifically, let $X_{i,con,obs}$, $X_{i,cat,obs},$ be the observed continuous and categorical variables of the $i$th observation.

Now we are ready to specify our model for mixed data. 

\begin{definition}[Latent Gaussian model for mixed data] Suppose that $W$ is a $p$-dimensional  Gaussian random vector with mean vector $\mu$ and covariance matrix $\Sigma$ and $V$ is a random vector of length $p$.
We say that $V \sim LGM(\mu, \Sigma)$ if 

(1)for any continuous component $j$, $V_j=W_j$, and

(2)for any categorical component $j$, $\mu_j=0$, $\sigma_{jj}=1$ and 
$$V_j=k \Longleftrightarrow W_j \in (C_{j,k},C_{j,k+1}),$$where $C_j$ is a fixed vector of thresholds of length $n_j+1$ with $n_j$ the level of categorical variable $V_j$.
\end{definition}
Note that generally, $C_{j,0} = -\infty$ and $C_{j,n_j} = \infty$.

If $X\in \mathbb R^{n\times p}$ are random samples of  $LGM(\mu, \Sigma)$, then from the definition, there exist $n$ observations $ Z_1, \dots ,Z_n $ from $N(\mu,\Sigma)$ such that  (1) for any continuous component $j$, $X_{ij}=Z_{ij}$, and
(2) for any categorical component $j$,
$$X_{ij}=k \Longleftrightarrow Z_{ij} \in (C_{j,k},C_{j,k+1}).$$

We call $Z_{ij}$ the latent value of $X_{ij}$. In the next section, we propose estimation method for the latent Gaussian model. We emphasize on missing data problems, which is more difficult than parameter estimation without missing values in  observed data $X$. Note that no matter for continuous variable or categorical variable (for example $X_j$), if one of the observation (for example $X_{ij}$) is missing the only information we have is that the latent value ($Z_{ij}$) lies in the interval $(-\infty, \infty)$. 
\section{Parameter Estimation}

Since there are a lot of parameters in this model. We use multi-stage method. First we estimate the collection of thresholds $C$ to determine categorical variables. Then we use the Expectation-Maximization algorithm to estimate the latent  mean and covariance matrix.  For high-dimensional data, covariance matrix might be sparse, finally we 
use graphical Lasso and constrained L1-minimization to obtain the sparse covariance matrix.

\subsection{Estimating the thresholds}
\label{ss:ETH}
The collection of thresholds $C$ is estimated first independently as they will be needed in the estimation of $\mu$ and $\Sigma$.

We only use the $j$th column of $X$ to compute the marginal likelihood of $C_{jk}$. It is not hard to show that the maximum marginal likelihood estimation is
\begin{equation}
\label{eqn:thr}
\hat{C}_{jk}=\Phi^{-1}\left(\frac{1}{n}\sum_{t=1}^{n}I(X_{tj}<k)\right), \mbox{ for } k = 0,1,\ldots, n_j,
\end{equation}
where $\Phi$ is the distribution function of the standard normal distribution. It is easy to check that $\hat C_{j0} = -\infty \mbox{ and } \hat C_{j,n_j} = \infty$.  It is obvious that $\hat{C}_{j,k}$ is a consistent estimate of $C_{j,k}$.

\subsection{Estimating Latent Gaussian Parameters}
\label{sub:ELGP}
We use the Expectation-Maximization algorithm to estimate $\mu$ and $\Sigma$ in the latent Gaussian model.
To apply the Expectation-Maximization algorithm we first need to compute the expectation of complete log likelihood given observed data and current estimation. As we have noted in Section 2, the complete data here is actually $Z$ (the latent value of $X$), and the observed data is $\{X_{i,con,obs}, X_{i,cat,obs}\}_{1\leq i\leq n}$. Suppose that the current parameter estimation is $\mu^{(t)}, \Sigma^{(t)}$. Since $Z_i$'s are independent and identically distributed samples from $N(\mu, \Sigma)$, the loglikelihood function of complete data is 
\begin{equation}
\label{llk}
\ell(\mu,\Sigma;Z)=C+\log (det(\Sigma)) +tr(S\Sigma^{-1})+ (\hat\mu-\mu)^T\Sigma^{-1}(\hat\mu-\mu),
\end{equation}
where $\hat \mu$ and $S$ is the sample mean and covariance matrix respectively. To compute the expected value of Equation \eqref{llk} given observed data we need to compute
\begin{equation}
\label{ConditionalCov}
E(Z_{ij}Z_{ik}|X_{i,con,obs}, X_{i,cat,obs}; \mu^{(t)}, \Sigma^{(t)}),
\end{equation}
for $1\leq i\leq n$ and $1\leq j,k \leq p$, and
\begin{equation}
\label{ConditionalMean}
E(Z_{ij}|X_{i,con,obs}, X_{i,cat,obs};\mu^{(t)}, \Sigma^{(t)}),
\end{equation}
for $1\leq i\leq n$ and $1\leq j \leq p$.
In the E-step, to estimate term \eqref{ConditionalCov} and \eqref{ConditionalMean}, we use the Gibbs sampling described in section \ref{subsub:e-steps}.  Once quantities  \eqref{ConditionalCov} and \eqref{ConditionalMean} are obtained, maximizing $E(\ell(\mu,\Sigma;Z)|X_{obs})$ over $\mu$ and $\Sigma$ is a standard procedure. The whole EM framework for mixed and missing data can now be stated as follows:
\begin{enumerate}
\item Start with random initial value $\mu^{(0)}, \Sigma^{(0)}$.
\item For $t=1,\dots$, $1\leq j,k \leq p$, let $\mu^{(t+1)}_j=\frac{1}{n}\sum_{i=1}^{n}E(Z_{ij}|X_{i,con,obs}, X_{i,cat,obs};\mu^{(t)}, \Sigma^{(t)})$ and $\Sigma^{(t+1)}_{jk}=\frac{1}{n}\sum_{i=1}^{n}E(Z_{ij}Z_{ik}|X_{i,con,obs}, X_{i,cat,obs};\mu^{(t)}, \Sigma^{(t)})-\mu^{(t+1)}_j\mu^{(t+1)}_k.$
\item Repeat (1) and (2) until convergence.
\end{enumerate}

\subsection{E-step}
\label{subsub:e-steps}

The previous subsection provides a framework for the Expectation-Maximization algorithm but did not show how to calculate the conditional expectations. We describe the calculations in detail in this subsection. 

We use Gibbs sampling to calculate the (conditional) expectations used in the EM algorithm.  Sampling from condition distributions given observed data in missing and mixed model could be viewed as  sampling from a truncated multivariate normal distribution.  This is because the complete data (provided that latent values are known) follows a joint multivariate normal distribution. For categorical observations, we know that latent values fall in some interval. Thus, the sampling could be view as sampling from the following conditional distribution:
$$f(W|a_j\leq W_j\leq b_j, j=1,\dots,p; \mu, \Sigma),$$
where $W$ is a $p$-dimensional  Gaussian random vector with mean vector $\mu$ and covariance matrix $\Sigma$; for missing values (no matter categorical or continuous), $a_j = -\infty$ and $b_j = \infty$; for continuous variables, $a_j = b_j$; for categorical variables, $a_j$ and $b_j$ are the thresholds that could be estimated via Equation \eqref{eqn:thr} in Section \ref{ss:ETH}.

Gibbs sampling from $f(W|a_j\leq W_j\leq b_j, j=1,\dots,p; \mu, \Sigma)$ could be done as follows. Starting from initial value $W^{(0)}=(W^{(0)}_1, \dots
W^{(0)}_p)$, for $t=1, \dots $ ,$j=1,\dots, p$ we need to sample $W^{(t)}_j$ from the one-dimensional conditional density
\begin{equation}
\label{onedimConditional}
\begin{aligned}
f(W^{(t)}_j|W^{(t)}_1,\dots, W^{(t)}_{j-1}, W^{(t-1)}_{j+1}, \dots, W^{(t-1)}_p, a_j<W^{(t)}_j<b_j; \mu, \Sigma)\\
\propto \exp\left(-\frac{(W^{(t)}_j-\mu^{(t)}_j)^2}{2(\sigma^{(t)}_j)^2}\right)I(a_j<W^{(t)}_j<b_j),
\end{aligned}
\end{equation}
where
\begin{equation}
\label{eqn:mean}
\mu^{(t)}_j=E(W_j|W^{(t)}_1,\dots, W^{(t)}_{j-1}, W^{(t-1)}_{j+1},\mu, \Sigma),
\end{equation}and 
\begin{equation}
\label{eqn:var}
(\sigma^{(t)}_j)^2=
Var(W_j|W^{(t)}_1,\dots, W^{(t)}_{j-1}, W^{(i-1)}_{j+1},\mu, \Sigma).
\end{equation}
Equations \eqref{eqn:mean} and \eqref{eqn:var} could be easily obtained since $W$ follows a joint multivatiate normal distribution with known parameters $\mu$ and  $\Sigma$.
We summarize the Gibbs sampling from $f(W|a_j\leq W_j\leq b_j, j=1,\dots,p; \mu, \Sigma)$ as in Algorithm \ref{alg:alg2}. These samples will be used to calculate conditional expectations as in Equations \eqref{ConditionalMean} and \eqref{ConditionalCov}.

\begin{algorithm}[h!]
\caption{Gibbs Sampling from truncated (conditional) normal distribution}
\label{alg:alg2}
\begin{algorithmic}
\STATE{Initialization. Set $W^{(0)}=(W^{(0)}_1, \dots
W^{(0)}_p)$.}
\WHILE{not converged}
\STATE Increase t by 1. t = t+1.
\FOR{$j = 1,2,\ldots,p$}
\STATE {Compute $\mu = \mu^{(t)}_j$ and $\sigma = \sigma^{(t)}_j$ according to Equations   \eqref{eqn:mean} and \eqref{eqn:var} .}
\STATE{Draw sample $u$ from the uniform distribution on $[0,1]$.}
\STATE {Compute $d=\Phi(\frac{a_j-\mu}{\sigma})$ and $e=\Phi(\frac{b_j-\mu}{\sigma})$.}
\STATE{Set $W^{(t)}_j=\mu+\sigma\Phi^{-1}(d+(e-d)u).$}
\ENDFOR
\ENDWHILE
\end{algorithmic}
\end{algorithm}

We use the following ergodicity result to show the validity of the Gibbs sampler. It is a special case of Lemma 3 in \cite{roberts1994simple}.
\begin{proposition}[Ergodicity]
\label{convergence}
For $W^{(i)}=\{W^{(i)}_1, \dots, W^{(i)}_p\}$, $i=1, \dots ,p$ generated from Algorithm \ref{alg:alg2}, 

(1) the density function of $W^{(i)}$ converges to $f(W|a_i<W_i<b_i, i=1,\dots,p;\mu, \Sigma)$;

(2) for any real-valued function $g$, if $E(g(W)|a_i<W_i<b_i, i=1,\dots,p;\mu, \Sigma)< \infty$, then as $N \rightarrow \infty$,
$$\frac{1}{N}\sum_{i=1}^{N}g(W^{(i)}) \rightarrow E(g(W)|a_i<W_i<b_i, i=1,\dots,p;\mu, \Sigma).$$
\end{proposition}

\subsection{Sparse inverse covariance selection}
The previous sections give a maximum likelihood estimation of parameters in the latent Gaussian model. But this is not enough, especially when the number of parameters is big. For such considerations, sparse inverse covariance selection is needed. Let $\hat\mu$, $\hat\Sigma$ be the maximum likelihood estimation of the mean and covariance matrix obtained by the Expectation-Maximization algorithm.Two methods could be used for sparse inverse covariance selection: graphical Lasso and constrained L1-minimization.

The Graphical lasso estimator is defined as
\begin{equation}
\Omega_G=\mathop{\arg\min}_{\Omega\geq0} tr(S\Omega)-\log(det(\Omega))+\lambda\sum_{i\neq j}|\Omega_{ij}|
\end{equation}
where $\lambda$ is the tuning parameter and $S$ is the sample covariance matrix. For the latent Gaussian model, we replace $S$ by $\hat{\Sigma}$ obtained via the likelihood-based method, and solve the following Graphical Lasso problem
$$
\Omega_G=\mathop{\arg\min}_{\Omega\geq0} tr(\hat{\Sigma}\Omega)-\log(det(\Omega))+\lambda\sum_{i\neq j}|\Omega_{ij}|.
$$
We refer the detail of Graphical Lasso and its algorithm to \cite{friedman2008glasso}.

We can also use constrained L1-minimization proposed by  \cite{cai2011clime} to get a sparse estimation of $\Sigma^{-1}$. The constrained L1-minimization estimation is defined as
\begin{equation}
\Omega_C=\mathop{\arg\min}_{\Omega} ||\Omega||_1, \quad s.t.\quad ||\hat{\Sigma}\Omega-I||_{max} \leq \lambda
\end{equation}
We refer the detail of this estimation and its related algorithm to \cite{cai2011clime}.

Using sparse inverse covariance selection (either Graphical Lasso or constrained L1-minimization), we obtain a family of covariance matrix $\Sigma_{\lambda}$ indexed by the tuning parameter $\lambda$. In classification problem, $\lambda$ is chosen such that the prediction error is minimized.

\section{Classification Using Latent Gaussian Model }
In this section we propose methods for classification using latent Gaussian model. We also point out that this model could be used for regression analysis and ranking problems. But for our purpose to deal with PCI data, we emphasize on classification problems. 

For mixed data without missing value, our model can be used for classification. When there are missing values, our model can be used for both classification and imputation.

We first discuss classification problem for mixed data when there is no missing value. Suppose that the $m$th variable (a categorical variable) is used to indicate which class one data point belongs to. We fit a latent Gaussian model for all variables. For a specific $\lambda$, denote $\Sigma_{\lambda}$ as the covariance matrix. For the $i$th observation in the test set, we compute $$p_k=P(X_{im}=k|X_{i,\setminus m};\mu, \Sigma_{\lambda})=P(Z_{im}\in(C_{m,k},C_{m,k+1})|X_{i,\setminus m};\mu, \Sigma_{\lambda})$$ via Gibbs sampling described in Section \ref{subsub:e-steps}. Then $X_{im}$ is assigned to category $k$ that maximizes $p_k$.  The tuning parameter $\lambda$ with the smallest prediction error is chosen.

When missing values exists, the previous principle could also be applied.  Denote by $\mu_{\lambda}, \Sigma_{\lambda}$ the parameters of the latent variables estimated by the training set, where $\lambda$ is the tuning parameter in graphical Lasso or constrained L1-minimization method. For the $i$th observation in the test set, we compute $$p_k=P(X_{im}=k|x_{i,obs\setminus m};\mu, \Sigma_{\lambda})=P(Z_{im}\in(C_{m,k},C_{m,k})|x_{i,obs\setminus m};\mu, \Sigma_{\lambda})$$ via Gibbs sampling. Likewise, $X_{im}$ is assigned to category $k$ that maximizes $p_k$ and the tuning parameter with the smallest prediction error is selected.

We could also use latent Gaussian model as an imputation model. That is, we impute the missing data in both the training set and the test set by sampling from latent Gaussian model. Then other traditional classification models, such as logistic regression and random forest, are used  to do classification.  We apply other methods instead of the latent Gaussian model (LGM) for classification purpose because the LGM might be misspecified, in which case, other methods could give better results. 

\section{Numerical Evaluation}
In this section, we evaluate the performance of our classification method using both numerical experiments and real data.
\subsection{Simulation}
\label{sub:simu}
We set $n=200,p=50$. To generate the inverse covariance matrix $\Omega$, we adopt the methods used by \cite{meinshausen2006glasso} and \cite{liu2012copula}. Set $\Omega_{jj}=1$, $j=1,2,...,n$ and $\Omega_{jk}=cr_{jk}$, $j\neq k$, where $c$ is a constant guaranteeing the positive definiteness of $\Omega$, and $r_jk$ are independent Bernoulli variables with $P(r_{jk}=1)=p_0\phi(d_{jk}/
\sqrt{p})$. $\phi$ is the density function of the standard normal distribution , $d_{jk}=||z_j-z_k||_2$, where $z_j$ are drawn from the uniform distribution on $[0,1]\times [0,1]$ and $p_0$ is a constant, controlling the sparsity of the true $\Omega$. In our experiment, we set $c=0.15$ and $p_0=1$.  Once  $\Omega$ is obtained, the covariance matrix $\Sigma$ = $\Omega^{-1}$.  We then rescale $\Sigma$ by taking $\Sigma_{ij} = \Sigma_{ij}/\sqrt{\Sigma_{ii}\Sigma_{jj}}$ so that each component of the Gaussian vector $X_j$ (j =1,2,\ldots,p) follows a standard normal distribution. For a categorical variable with $k$ levels, the $k-1$ non-trivial thresholds (the first and the last ones are $-\infty$ and $+\infty$ respectively) are drawn independently from the uniform distribution on [-1,1].\par
We consider the following four data generating scenarios. In each scenario, we first draw $W \in \mathbb R^{n\times p},$ with each row i.i.d. from $\sim N(0,\Sigma)$ and then cut a few continuous variable into categorical variables. Each point $C_{jk}$ in the following settings is drawn independently from the uniform distribution on [-1,1].
\begin{enumerate}
\item Generate $n$ observations with $49$ continuous variables and one categorical (binary) variable.  For $j = 1,2\ldots, 49$, let $X_j = W_j$  and $X_{50}=I(W_{50}>C_{50,1}$).
\item Generate $n$ observations with $20$ continuous variables and $30$ categorical (binary) variables.  For $j = 1,2\ldots, 20$, let $X_j = W_j$  and for $j=21,22,\ldots,50$, $X_{j}=I(W_{j}> C_{j1}$).  
\item Generate $n$ observations with 20 Gaussian variables, 10 binary variables ,10 categorical variables with 3 levels and 10 categorical variables with 4 levels.  Set $X_j=W_j$ for $j=1, \dots,20$,$X_j=I(W_j>C_{j1})$ for $j=21,...,30$, $X_j=I(W_j>C_{j1})+I(Z_j>C_{j2})$ for $j=31,...,40$, $V_j=I(W_j>C_{j1})+I(W_j>C_{j2})+I(W_j>C_{j3})$ for $j=41,...,50$.
\item Generate $n$ observations with categorical random variables only, of which 20  are binary variables ,10 are  variables with 3 levels and 20 are variables with 4 levels.
 Set $X_j=I(W_j>C_{j1})$ for $j=1,...,20$, $X_j=I(W_j>C_{j1})+I(W_j>C_{j2})$ for $j=21,...,30$, $V_j=I(W_j>C_{j2})+I(W_j>C_{j2})+I(W_j>C_{j3})$ for $j=31,...,50$.
\end{enumerate}

Now we generate missing values in each of the four settings. We use a missing at random mechanism. Ten variables are selected at random. For each observation and each selected variable, we decide whether the value is missing or not. We use $Rest$ to denote the other 40 variables that are observed completely. We use $R_{ij}$ to denote the indicator of missingness, i.e. $R_{ij}= 1$ if $X_{ij}$ is missing, and $R_{ij}= 0$ otherwise. We model $R_{ij}$ using a logistic regression. The probability that $X_{ij}$ is missing is given by
\[P(R_{ij}=1) = \frac{\exp\{-1+\beta\sum_{k\in Rest}x_{ik}\}}{1 + \exp\{-1+\beta\sum_{k\in Rest}x_{ik}\}},\]
where the coefficient $\beta$ is tuned such that the missing proportion for each variable is around $40\%$. This trick has been used in \cite{garcia2010variable}.

We use the EM algorithm described in Section  \ref{sub:ELGP} to find matrix $\Sigma$. Then we use graphical Lasso and constrained L1-minimization method to estimate sparse concentration matrix $\Omega$ . 

In Table \ref{simulationgraphical}, we list the Frobenius norm and spectral norm of the matrix $\hat{\Omega}-\Omega$ in different scenarios. The tuning parameter is chosen to be the oracle parameter which minimizes the two norms mentioned above.

\begin{table}[h!]
\centering
\begin{tabular}{lccc}
\hline
&&Graphical Lasso&constrained L1-minimization\\ \hline
(1)&F&3.20(3.43)&3.21(3.47)\\
&S&1.20(1.27)&1.30(1.37)\\ \hline
(2)&F&3.96(4.23)&4.10(4.35)\\
&S&2.21(2.27)&2.25(2.31) \\ \hline
(3)&F&3.79(4.04)&3.86(4.10)\\
&S&2.09(2.17)&2.17(2.25)\\ \hline
(4)&F&3.92(4.19)&4.05(4.28)\\
&S&2.16(2.24)&2.23(2.31)\\ \hline
\end{tabular}
\caption{The Frobenius(F) and Spectral(S) norm of $\hat{\Omega}-\Omega$ in different data generating scenarios. Given in parenthesis are the corresponding norms when $X$ has missing values.}
\label{simulationgraphical}
\end{table}


It can be seen from Table \ref{simulationgraphical} that 
the  loss decreases as the number of category increases. For example, all 30 categorical variables in scenario 2 are binary variables whereas 20 categorical variables in scenario 3 have more than 2 categories, and the estimation of $\Omega$ is better in scenario 3. This observation matches our intuition since a categorical variable with more levels approximates the latent variable better than one with less levels. 

Now we examine our model's performance for classification. For complete data sets, we compare the prediction accuracy of our model with random forest, which is known to be the state-of-art method for classification problems. Let the outcome variable be the binary variable with the smallest absolute threshold value. We generate 100 data sets in each scenario and use 5-fold cross validation to compute the prediction error. The result is shown in Table \ref{simulationcomplete}.

\begin{table}[h!]
\centering
\begin{tabular}{ccc}
\hline
&Latent Gaussian model&Random forest\\ \hline
(1)&32.4\%(7.7\%)&34.4\%(6.3\%)\\ 
(2)&35.9\%(8.2\%)&37.2\%(10.2\%)\\
(3)&34.0\%(8.1\%)&36.9\%(7.5\%)\\
(4)&39.3\%(7.2\%)&38.5\%(7.0\%)\\
\hline
\end{tabular}
\caption{Prediction error for different models with complete data. Standard deviations are given in parenthesis.}
\label{simulationcomplete}
\end{table}

For data sets with missingness, our model can be used for both classification and imputation. The following three methods will be used here fro comparasion.
\begin{enumerate}
\item Use latent Gaussian model as classification model. 
\item Use latent Gaussian model as imputation model, and random forest as classification model. 
The tuning parameter with the smallest prediction error is selected and the corresponding prediction error is reported.
\item   Use hot deck as imputation model, and random forest as classification model. This method is included as a benchmark for the prediction error. \end{enumerate}

We generate 100 data sets under each scenario and use 5-fold cross-validation to estimate the prediction error. For 2 and 3, we impute the missing data 100 times and calculate the mean (but not the variance) of the prediction error. The result is shown in Table \ref{simulationmissing}.
\begin{table}
\centering
\begin{tabular}{cccc}
\hline
&Latent Gaussian model&Latent Gaussian model+Random forest&Hot deck+Random forest\\ \hline
(1)&33.1\%(8.5\%)&35.1\%(10.5\%)&35.9\%(9.7\%)\\
(2)&36.6\%(9.1\%)&37.6\%(11.5\%)&37.7\%(11.2\%)\\
(3)&35.3\%(8.8\%)&37.8\%(11.0\%)&37.4\%(10.6\%)\\
(4)&39.9\%(7.9\%)&39.1\%(10.7\%)&39.7\%(9.9\%)\\
 \hline
\end{tabular}
\caption{Prediction error for different models with missing data. Standard deviations(calculated across data sets)are given in parenthesis.}
\label{simulationmissing}
\end{table}

It can be seen that our LGM model performs better than random Forest in most cases. Only when all variables are categorical (case 4 in Table \ref{simulationmissing}), LGM performs slightly worse. The result matches our intuition since when the proportion of continuous variables are higher, data contains more information on $\mu$ and $\Sigma$. In the case of is missing data, the latent Gaussian model also proves to be a better imputation model than naive imputation methods.

\subsection{Analysis of PCI (percutaneous coronary intervention) data}
In this section, we consider the PCI (percutaneous coronary intervention) data provided by two hospitals in Beijing. The data consists of 2582 observations. For each observation, there are 22 categorical variables and 16 continuous variables that are measured before the operation and a binary variable indicating whether there is reflux after PCI. All 39 variables have missing data with missing proportion ranging from less than 1\% to over 70\%.  We use our proposed method to determine the graphical model of the variables and to predict the outcome of the operation given the 38 variables measured before the operation.

\begin{figure}[h!]
\centering
\includegraphics[width=0.5\textwidth]{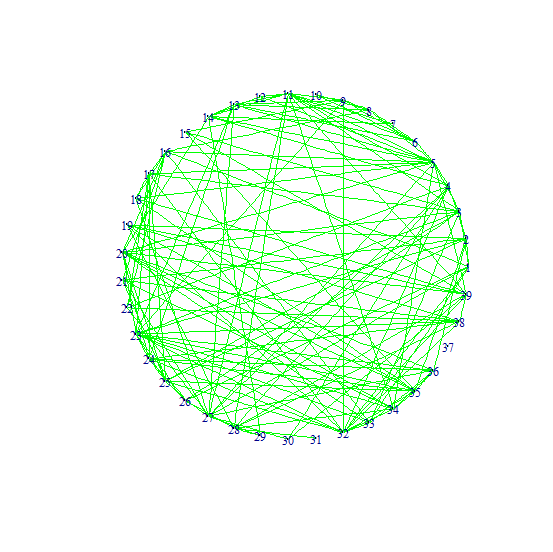}
\caption{The graphical model with the tuning parameter that minimizes the prediction error. The 39th variable is the binary response.}
\label{graph}
\end{figure}

First, we fit a latent Gaussian model with all observations using the Expectation-Maximazation algorithm, as described in Section \ref{sub:ELGP}. Then we use graphical Lasso to obtain a family of sparse inverse covariance matrix indexed by the tuning parameter $\lambda$.
In Table \ref{PCIgraphicalmodel}, we listed the number of nonzero elements in the last column of the inverse covariance matrix and the new variable entering $\Omega$ under different $\lambda$. We know that when $\lambda$ is big, all elements of $\hat \Omega$ will be zeros. When we decrease $\lambda$, some variable will enter into the estimation.  We monitor the last column of $\hat \Omega$. Since the nonzero elements in this column mean that these variables have prediction power for the outcome.  The first three non-zero entries that enter into  the last column of the inverse covariance matrix correspond to the variables hypertension history, blood glucose and AP(angina pectoris). The association of these variables with the outcome of PCI was revealed both from a medical perspective and by other statistical methods (such as variable selection of logistic regression), as in \cite{madani2013elective} and \cite{dumont2006predictors}. As an illustration of the latent sparse inverse covariance matrix, we plot in Figure \ref{graph} the latent Gaussian graphical model with the tuning parameter that minimizes the test prediction error.
\begin{table}[h!]
\centering
\begin{tabular}{lcc}
\hline
$\lambda$&Number of nonzero elements&New variable\\ \hline
0.13&1&Hypertension history\\
0.12&2&Blood glucose\\
0.10&3&AP(angina pectoris)\\
0.09&4&CRP\\
0.08&5&Admitting diagnosis\\
 \hline
\end{tabular}
\caption{The number of nonzero elements in the last column of the inverse covariance matrix and the new variable entering the last column of $\Omega$ with decreasing $\lambda$.}
\label{PCIgraphicalmodel}
\end{table}

We now turn to predict the binary outcome of the operation. The same methods 1-3 (1. pure latent gaussian model; 2. latent gaussian model + random forest/logistic; 3. hot deck imputation + random forest/logistic) as in section \ref{sub:simu} are applied, but random forest and logistic regression are both used here for complete data inference (i.e. inference after imputation). Similarly, we use 5-fold cross-validation to estimate the prediction error. For 2 and 3, we impute the missing data 100 times and calculate the mean of the prediction error. The tuning parameter with the smallest prediction error is selected and the corresponding prediction error is reported. The result is presented in Table \ref{PCIerror}. Here the latent Gaussian model is potentially (and almost certainly) misspecified. Nevertheless, we get the smallest prediction error when we use latent Gaussian model as the imputation model and random forest as the classification model.

\begin{table}[h!]
\begin{center}
\begin{tabular}{cc}
\hline
Method&Prediction Error\\ \hline
Latent Gaussian model&16.2\%\\
Latent Gaussian model+Random forest&13.0\%\\
Hot deck+Random forest&13.9\%\\
Latent Gaussian model+logistic&20.9\%\\
Hot deck+logistic&21.0\%\\
\hline
\end{tabular}
\caption{Prediction error of different methods for the PCI data.}
\label{PCIerror}
\end{center}
\end{table}


\section*{Acknowledgement}
This work is partially supported by the National Basic Research Program of China (973 Program 2011CB809105), NSFC-61121002, NSFC-11101005, DPHEC-20110001120113, and MSRA.
We are grateful to Doctor Jinwen Wang for providing PCI data.

\bibliographystyle{biometrika}
\bibliography{MGMreferences.bib}
\end{document}